\documentclass{osa-article}

\journal{osajournal}

\articletype{Research Article}

\begin{document}

\title{Octave-spanning microcomb generation in
4H-silicon-carbide-on-insulator photonics platform}

\author{Lutong Cai,\authormark{1} Jingwei Li,\authormark{1}, Ruixuan Wang \authormark{1} and Qing Li\authormark{*1}}

\address{\authormark{1}Department of Electrical and Computer Engineering, Carnegie Mellon University, Pittsburgh, PA 15213, USA}

\email{\authormark{*}qingli2@andrew.cmu.edu} 


\begin{abstract}
    Silicon carbide has recently emerged as a promising photonics material due to its unique properties, including possessing strong second- and third-order nonlinear coefficients and hosting various color centers that can be utilized for a wealth of quantum applications. Here, we report the design and demonstration of octave-spanning microcombs in a 4H-silicon-carbide-on-insulator platform for the first time. Such broadband operation is enabled by optimized nanofabrication achieving >1 million intrinsic quality factors in a 36-$\mu$m-radius microring resonator, and careful dispersion engineering by investigating the dispersion properties of different mode families. For example, for the fundamental transverse-electric mode whose dispersion can be tailored by simply varying the microring waveguide width, we realized a microcomb spectrum covering the wavelength range from 1100 nm to 2400 nm with an on-chip power near 120 mW. While the observed comb state is verified to be chaotic and not soliton, attaining such a large bandwidth is a crucial step towards realizing $f$-2$f$ self-referencing. In addition, we have also observed coherent soliton-crystal state for the fundamental transverse-magnetic mode, which exhibits stronger dispersion than the fundamental transverse-electric mode and hence a narrower bandwidth.
\end{abstract}

\section{Introduction}
An optical frequency comb is a coherent light source that has applications across a multitude of areas including frequency generation and synthesis \cite{Papp_comb_synthesizer, Comb_freq_XK, Wong_SiN_THz, Diddams_review_spectrum}, imaging and sensing \cite{Vahala_comb_imaging, Diddams_comb_sensing1, Comb_midIR_spectroscopy, Vahala19_comb_exoplanets}, light detection and ranging \cite{Vahala_comb_ranging, Kippenberg_comb_ranging}, parallel optical communication and computation \cite{Comb_communication, Comb_parallel_computation}, and quantum information processing \cite{Wong_SiN_quantum, Review_quantum_comb}. The last decade has seen the development of on-chip comb sources implemented in micro-resonators (i.e., microcombs) featuring low power consumption, compact form, and excellent phase stability \cite{Diddams_comb_review1, Gaeta_comb_battery, Bowers_comb_integration, Bowers_turnkey_soliton, Bowers_comb_laser}. So far, microcombs have been realized in a variety of integrated photonic platforms such as silicon \cite{Gaeta_Si_comb, Gaeta_Si_dual_midIR}, silica \cite{Vahala_comb_silica, Xiao_comb_silica}, aluminum nitride \cite{Guo_comb_AlN, Tang_comb_AlN_ref}, silicon nitride \cite{Gaeta_SiN_octave, Li_SiN_octave, Kippenberg_SiN_octave}, lithium niobate \cite{Loncar_LN_EOM, Lin_LN_soliton, Tang_LN_comb}, AlGaAs \cite{Pu_AlGAAs_comb, Bowers_comb_AlGaAs}, etc. Recently, silicon carbide (SiC) has emerged as a promising photonics material due to its unique material properties including a wide transparency window (approximately 400-5500 nm), simultaneously possessing strong second- and third-order optical nonlinearities, a relatively high refractive index (nominal index around 2.6 in the telecommunication band), and a large thermal conductivity coefficient. In addition, SiC hosts a variety of color centers that can be exploited for vital quantum devices such as single-photon sources and quantum memories \cite{Awschalom_SiC_qubit, Awschalom_SiC2}. The combination of these properties renders SiC an attractive platform for photonic applications in both the classical and quantum domains \cite{Vuckovic_SiC_review}.

In this work, we are interested in generating broadband microcombs based on the strong third-order nonlinear coefficient (i.e., Kerr combs) offered by SiC \cite{SiC_nonlinear_coeff, Lin_3CSiC_nonlinear, Gaeta_4HSiC_nonlinear}. Thanks to the direct wafer bonding and polishing technique, thin-film SiC-on-insulator (SiCOI) platforms have been successfully demonstrated for various polytypes of SiC, with the two most common choices being 3C \cite{Lin_3CSiC, Adibi_3CSiC, Adibi_3CSiC_Q} and 4H \cite{Gaeta_4HSiC_nonlinear, Ou_4HSiC, Noda_4HSiC_PhC, Vuckovic_4HSiC_nphoton, Ou_4HSiC_combQ}. To date, microresonators implemented in 4H-SiCOI have achieved the highest quality factors ($Q$s), with the state-of-the-art reaching the 5-7 million range \cite{Vuckovic_4HSiC_soliton, Ou_4HSiC_combQ}. In addition, several important applications such as single-photon emission \cite{Awschalom_SiC_PhC}, optical parametric oscillation \cite{Vuckovic_4HSiC_MIcomb}, optomechanical resonances \cite{Lin_3CSiC_double}, and harmonic frequency generations \cite{Noda_4HSiC_PhC, Vuckovic_4HSiC_nphoton, Ou_4HSiC_combQ} have been investigated. Only recently, microcomb generation was reported in 4H-SiCOI with limited performance \cite{Vuckovic_4HSiC_MIcomb, Vuckovic_4HSiC_soliton, Ou_4HSiC_combQ}. For instance, an incoherent modulation-instability comb state was demonstrated in a high-$Q$ SiC microdisk, achieving the largest wavelength span so far from 1300 nm to 1700 nm \cite{Ou_4HSiC_combQ}. In another work, coherent soliton and soliton-crystal states were reported, though the wavelength bandwidth was estimated to be less than 200 nm \cite{Vuckovic_4HSiC_soliton}. For many metrology and timekeeping related applications, however, it is necessary to obtain a comb bandwidth on the order of one octave or more \cite{Papp_comb_synthesizer, Diddams_review_spectrum}. Reaching such a technological milestone is especially meaningful given that 4H-SiC has strong second-order nonlinearity which can be used to double the frequency through the second-harmonic generation \cite{Noda_4HSiC_PhC, Vuckovic_4HSiC_nphoton}, thus enabling the $f$-2$f$ self-referencing process on the same chip \cite{Diddams_comb_selfref, Gaeta_LN_selfref, Tang_comb_AlN_ref}.

\begin{figure}[h]
\centering
\includegraphics[width=0.95\linewidth]{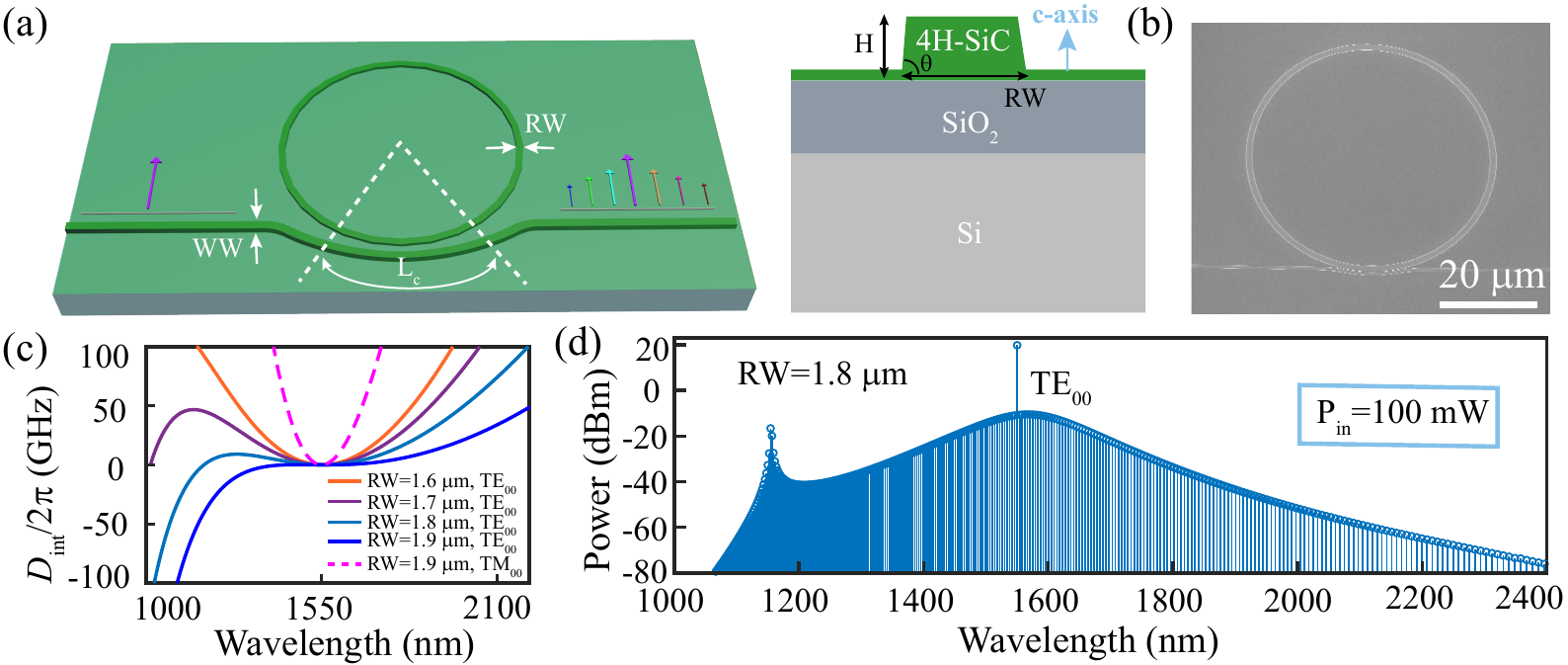}
\caption{(a) Schematic top view (left) and cross-section (right) of the 4H-silicon-carbide-on-insulator platform for frequency comb generation based on compact microring resonators. The sidewall angle ($\theta$) is estimated near 80-85 degrees in our nanofabrication. Dispersion engineering is carried out by varying the ring waveguide width (RW). In addition, efficient coupling is realized using the pulley structure where the access waveguide width and coupling length are adjusted to achieve phase matching to the desired resonant mode families. (b) Scanning electron micrograph of a 36-$\mu$m-radius SiC microring. In this work the SiC thickness is fixed at 500 nm with a pedestal layer of 100 nm. (c) Simulated integrated dispersion ($D_\text{int}$, see its definition in Eq.~\ref{Eq1}) for the fundamental transverse-electric (i.e., TE$_{00}$) mode with ring widths varied from 1.6 $\mu$m to 1.9 $\mu$m (solid lines). For comparison, $D_\text{int}$ for the fundamental transverse-magnetic (i.e., TM$_{00}$) mode is also provided for the ring width of 1.9 $\mu$m (magenta dashed line), exhibiting much larger values than those of the TE$_{00}$ modes. (d) Simulated comb spectrum for the TE$_{00}$ mode for a 36-$\mu$m-radius SiC microring with RW $=1.8$ $\mu$m and an input power of 100 mW, featuring a spectral bandwidth of more than one octave and dispersive-wave generation near the wavelength of 1150 nm. In the simulation, the Kerr nonlinear parameter is assumed to be $\gamma \approx 2.1\ \text{W}^{-1}\text{m}^{-1}$, and the intrinsic and loaded quality factors are assumed to be 1.25 million and 0.75 million, respectively.}
\label{Fig_Schematic}
\end{figure}
Here, we report the first octave-spanning microcomb generation in the 4H-SiCOI platform. This is achieved by engineering the dispersion of the fundamental transverse-electric (TE$_{00}$) mode in a compact SiC microring resonator (36-$\mu$m radius) on the one hand, and optimizing the nanofabrication to obtain intrinsic $Q$s above 1 million on the other hand. With these efforts, we realized a microcomb with wavelength coverage from 1100 nm to 2400 nm with approximately 120 mW on-chip power. While the observed comb state is verified to be chaotic, attaining such a large bandwidth is a crucial step towards realizing $f$-2$f$ self-referencing. In addition, we have also observed coherent soliton-crystal state for the fundamental transverse-magnetic (TM$_{00}$) mode, which exhibits stronger dispersion than the TE$_{00}$ mode and hence a narrower bandwidth. 

\section{System design and overview}
The goal of this work is to develop SiC device technologies that enable broadband combs with a telecommunication C-band pump laser (i.e., near 1550 nm). As illustrated in Fig.~\ref{Fig_Schematic}(a), our device platform is based on 4H-SiCOI, where the under-cladding silicon dioxide (SiO$_2$) layer is set to be approximately 2 $\mu$m to provide sufficient isolation from the silicon substrate. To attain a large comb bandwidth, dispersion properties for various resonant modes of a 36-$\mu$m-radius SiC microring are investigated \cite{Kim_comb_dispersion, Li_SiN_octave}. Here we fix the SiC thickness to be 500 nm while maintaining an approximately 100-nm-thick pedestal layer. Our optimized nanofabrication yields accurate dimensional control in the waveguide width; in particular, the sidewall angle $\theta$ defined in Fig.~1(a) is estimated to be larger than 80 degrees. This enables a straightforward implementation of the pulley coupling structure depicted in Fig.~1(a), where selective coupling to the desired resonant mode family is achieved by employing the phase-matched access waveguide width and an appropriate coupling length \cite{Li_FWMBS}. Figure 1(b) shows the scanning electron micrograph (SEM) of a 36-$\mu$m-radius microring used in this work, which supports resonances with $>1$ million intrinsic $Q$s (details in Sect.~3)

By computing the resonance frequencies in an eigenmode solver (COMSOL Multiphysics), we extract the group-velocity dispersion (GVD) and higher-order dispersion terms for different ring waveguide widths. For example, Fig.~1(c) plots the simulated integrated dispersion ($D_\text{int}$) for the TE$_{00}$ mode, which is defined as:
\begin{equation}
D_\text{int} \equiv \omega_{\mu} - \omega_0 - D_1\mu = \sum_{k=2}^{\infty} \frac{1}{k!}D_k\mu^k,
\label{Eq1}
\end{equation}
where $\mu$ is the relative azimuthal order to the pump resonance (i.e., $\mu=0$ for the pump mode); $\omega_{\mu}$ is the corresponding resonance frequency; $D_1$ is the free spectral range (FSR) of the resonator; and $D_2$ and $D_k$ ($k>2$) represent the GVD and  $k_\text{th}$-order dispersion, respectively \cite{Kippenberg_soliton}. The GVD of the TE$_{00}$ mode can be tuned by simply varying the ring waveguide width (as can be seen in Fig.~1(c) by focusing on the wavelength region near the pump resonance where $D_2$ is the dominant term), changing from relatively strong anomalous dispersion for $\text{RW}=1.6\ \mu$m to weakly anomalous for $\text{RW}=1.9\ \mu$m (the GVD of the TE$_{00}$ mode becomes normal for $\text{RW}>2.0\ \mu$m). In comparison, the GVD for the TM$_{00}$ mode is much stronger and not very sensitive to the waveguide width tuning. Hence, we focus on the TE$_{00}$ mode family for broadband comb generation. 

To reach an octave span in the comb spectrum, higher-order dispersion terms need to be included in the design. The integrated dispersion profile displayed in Fig.~1(c) suggests that the comb bandwidth can be further broadened by the so-called dispersive-wave generation at shorter wavelengths of the pump laser for ring widths above $1.7\ \mu$m \cite{Kippenberg_DW}. This is confirmed by numerical simulation based on the Lugiato-Lefever equation (LLE) that takes high-order dispersion terms into consideration \cite{Coen_LLE}. The simulation example provided in Fig.~1(d) is for $\text{RW}=1.8\ \mu$m with a pump power of 100 mW, where an octave-spanning comb bandwidth has been realized. In addition, a dispersive wave is generated near the wavelength of 1150 nm, in agreement with the dispersion data in Fig.~1(c) (the position of the dispersive wave is located around $D_\text{int}\approx 0$ for nonzero $\mu$) \cite{Coen_LLE, Li_SiN_octave}. 

\section{Fabrication and linear characterization}
The device fabrication begins by preparing a 4-inch 4H-SiCOI wafer using the typical bonding and polishing process \cite{Noda_4HSiC_PhC, Vuckovic_4HSiC_nphoton, Ou_4HSiC_combQ}: first, a 2-$\mu$m-thick SiO$_2$ layer is deposited on an undoped 4H-SiC wafer (STMicroelectronics Silicon Carbide AB in Sweden) based on plasma-enhanced chemical vapor deposition (PECVD). After bonding SiO$_2$ to a silicon handle wafer, the SiC layer is thinned down from the initial thickness of $500\ \mu$m to $<1\ \mu$m through the grinding and chemical and mechanical polishing (NGK Insulators). We then dice the 4H-SiCOI wafer into 1 cm $\times$ 1 cm chips and further thin down the SiC layer to the desired thickness depending on the application. This is achieved by running a CHF$_3$/O$_2$ plasma etching process in a reactive-ion etching system (Plasma Therm). The etch rate is slow enough ($\approx$ 7.25 nm/min) to allow accurate control of the final SiC thickness. Next, the nanofabrication proceeds by spin-coating 1-$\mu$m-thick negative e-beam resist (flowable oxide, FOx-16) as the etching mask and the device patterns are defined using a 100-kV electron-beam lithography system (Elionix ELS-G100). After development, the pattern is transferred to SiC using the same CHF$_3$/O$_2$ plasma etching process that is used to fine tune the SiC thickness. The inset in Fig.~2(a) shows the SEM picture of an etched SiC waveguide, displaying smooth sidewalls which are expected to minimize the scattering loss on the surface and result in high-$Q$ optical resonances.

\begin{figure}[h]
\centering
\includegraphics[width=0.9\linewidth]{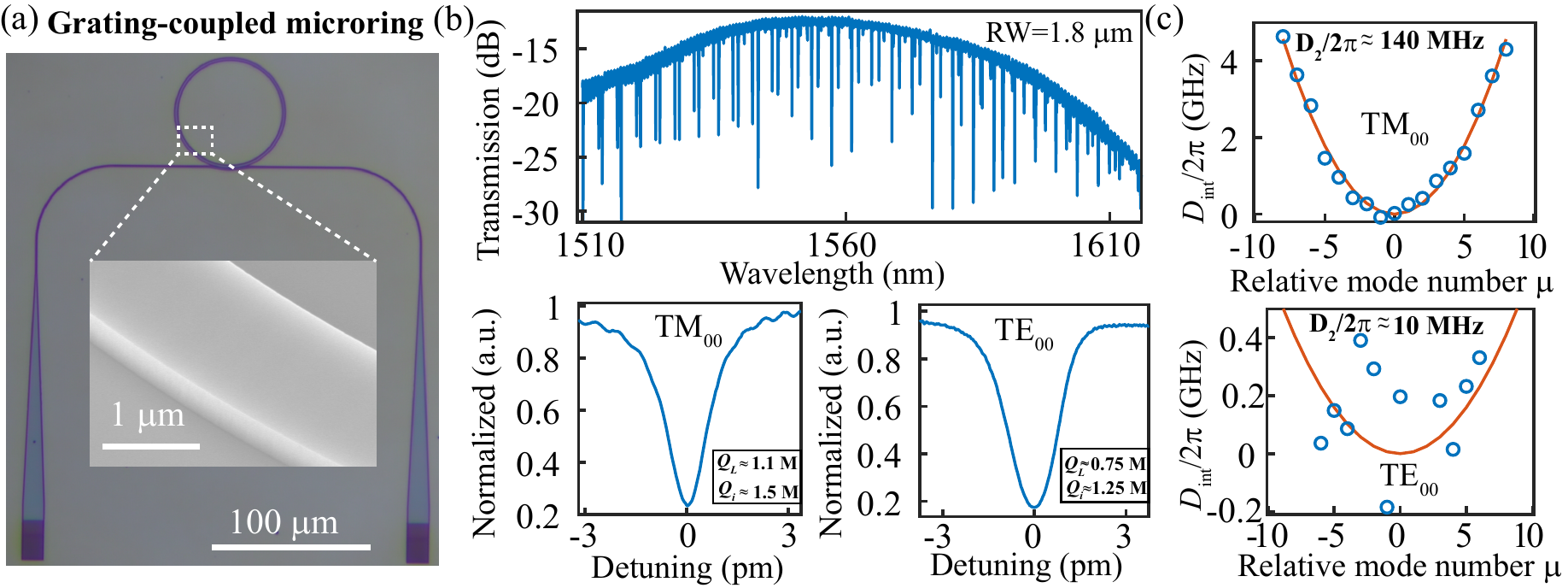}
\caption{(a) Optical micrograph of a 36-$\mu$m-radius SiC microring with grating input and output. The inset shows the scanning electron micrograph for one section of the ring waveguide, which exhibits smooth sidewalls and $>80$ degree sidewall angles. The access waveguide widths are set to be near 800 nm and 1000 nm for the efficient coupling to the TM$_{00}$ and TE$_{00}$ modes, respectively. (b) (top) Exemplary transmission of a grating-coupled microring resonator, with a typical insertion loss around 5-7 dB per grating and a 3-dB coupling bandwidth around 60-80 nm. The grating period is optimized to center the peak efficiency near the wavelength of 1550 nm; (bottom) Representative resonances with normalized transmission for the TM$_{00}$ and TE$_{00}$ modes, both with intrinsic $Q$s above 1 million ($Q_L$ and $Q_i$ stand for the loaded $Q$ and intrinsic $Q$, respectively). (c) Dispersion characterization of the TM$_{00}$ and TE$_{00}$ mode families, displaying an estimated second-order dispersion ($D_2/2\pi$) near 140 MHz and 10 MHz, respectively.}
\label{Figure_linear}
\end{figure}
In linear characterization, the light emitted from a telecommunication-band tunable laser (Agilent 81642A with linewidth <100 kHz) is coupled to the on-chip waveguide from a fiber V-groove array (VGA) based on a grating coupler, with its period optimized for either the TE- or TM-polarized light (see Fig.~2(a) for the optical micrograph). Figure 2(b) shows a typical swept-wavelength transmission for a 36-$\mu$m-radius SiC microring, where the output light in the waveguide is coupled back to the VGA (a different fiber though) by a similar grating coupler as the input. The insertion loss from each grating is estimated to be 5 to 7 dB in the center, while its 3-dB coupling bandwidth is limited to 60-80 nm due to the dispersive nature of the grating. By comparing the measured FSRs to theoretical values, we can identify the mode family of each resonance, and critical information such as $Q$s and GVDs is extracted \cite{Li_FWMBS}. In Fig.~2(b), two representative resonances are plotted for the TM$_{00}$ and TE$_{00}$ modes near 1550 nm, exhibiting intrinsic $Q$s around 1.5 million and 1.25 million, respectively. The GVDs of these two modes are also verified to be close to the simulation data provided in Fig.~1(c). For instance, for the microring with $\text{RW}=1.8\ \mu$m, we extract its second-order dispersion ($D_2/2\pi$) to be near 140 MHz and 10 MHz for the TM$_{00}$ and TE$_{00}$ mode families, respectively (Fig.~2(c)).   

\section{Microcomb generation}
\begin{figure}[h!]
\centering
\includegraphics[width=0.9\linewidth]{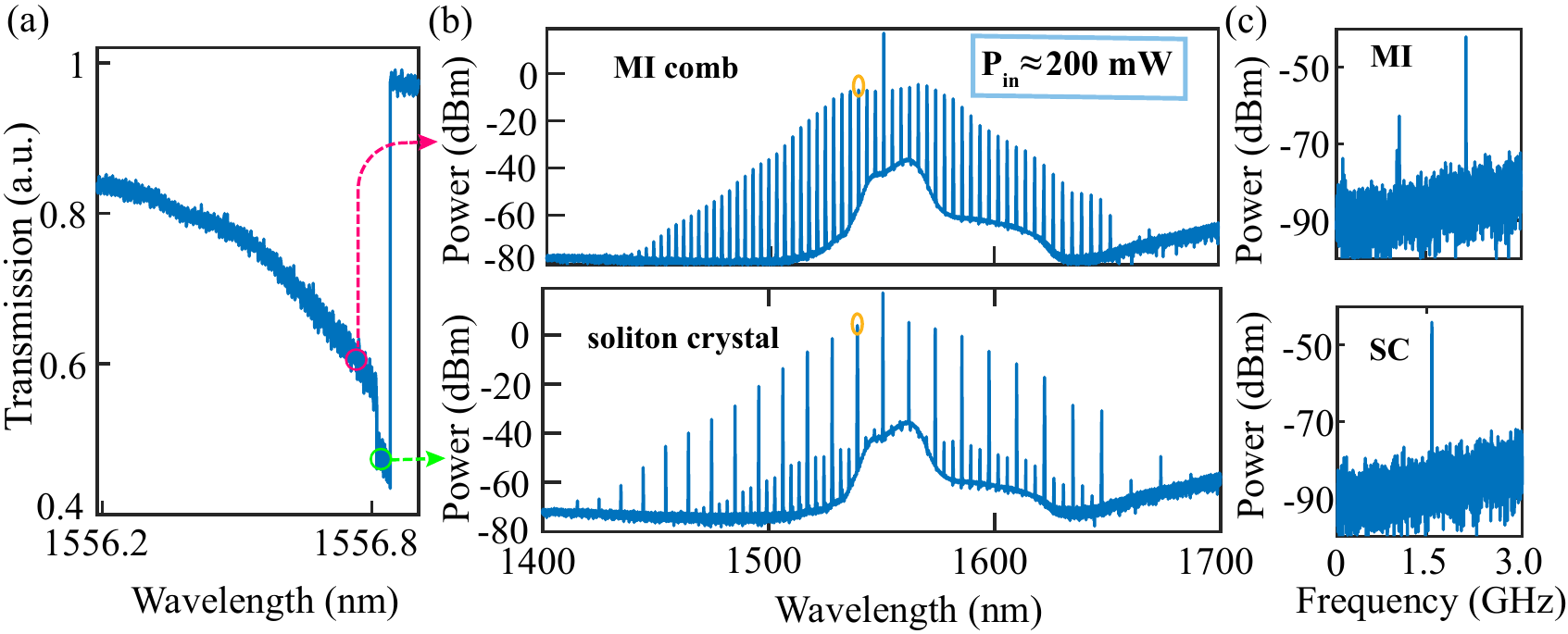}
\caption{(a) Nonlinear transmission of the TM$_{00}$ mode with an input power near 200 mW. The resonance is broadened due to the thermo-optic and the Kerr effects. The SiC microring has a width of $2.0\ \mu$m. (b) Comb spectra corresponding to the chaotic modulation-instability (MI) state and the coherent soliton-crystal (SC) state. These two different states are generated by slowly varying the laser detunings, with their respective positions marked in (a). (c) Coherence characterization by filtering out one of the comb lines (marked by circles in (b)) and measuring its beat-note signal with an auxiliary narrow-linewidth laser tuned into similar wavelengths: (top) the MI comb state has shown multiple RF peaks while (bottom) the SC state has shown a single RF peak which shifts with the wavelength of the auxiliary tunable laser.}
\label{Figure_TM_comb}
\end{figure}

To generate frequency combs, the laser output is amplified by an erbium-doped fiber amplifier (EDFA, maximum power up to 2W) before being coupled to the on-chip waveguide using a grating coupler. For relatively narrowband combs based on TM$_{00}$ modes, we out-couple the light using another grating coupler with the same configuration as shown in Fig.~2(a). An exemplary nonlinear transmission for the TM$_{00}$ mode of a 36-$\mu$m-radius SiC microring ( $\text{RW}=2.0\ \mu$m) is provided in Fig.~3(a), where the laser wavelength is increased from the initial blue-detuned regime to the final red-detuned regime at a slow speed. As shown, the nonlinear resonance is significantly broadened by the thermo-optic (TO) and Kerr effects. In addition, as we vary the pump laser detunings, different comb states including the modulation-instability (MI) comb and soliton-crystal (SC) state are observed (Fig.~3(b)) \cite{Kippenberg_soliton, Papp_soliton_crystal}. The MI comb features one-FSR spacing in the spectral domain while exhibiting chaotic behavior in the time domain. Its coherence property is characterized by filtering out one of the comb lines (marked by circles in Fig.~3(b)) and measuring its beat-note signal with an auxiliary laser tuned into similar wavelengths \cite{Li_SiN_octave}. The multiple peaks in the corresponding radio-frequency (RF) spectrum (Fig.~3(c)) indicate that the MI comb is indeed incoherent. On the other hand, the SC state observed in Fig.~3(b) has a spacing of three FSRs and features a single RF peak whose position shifts with the wavelength of the auxiliary laser, a strong sign that this comb state is coherent \cite{Kippenberg_soliton}. Finally, we note that the approximate 250 nm wavelength span in both the MI and SC comb states is partially limited by the out-coupling bandwidth of the grating coupler, which has a 3-dB bandwidth less than 100 nm (see Fig.~2(b)). 

\begin{figure}[h!]
\centering
\includegraphics[width=0.85\linewidth]{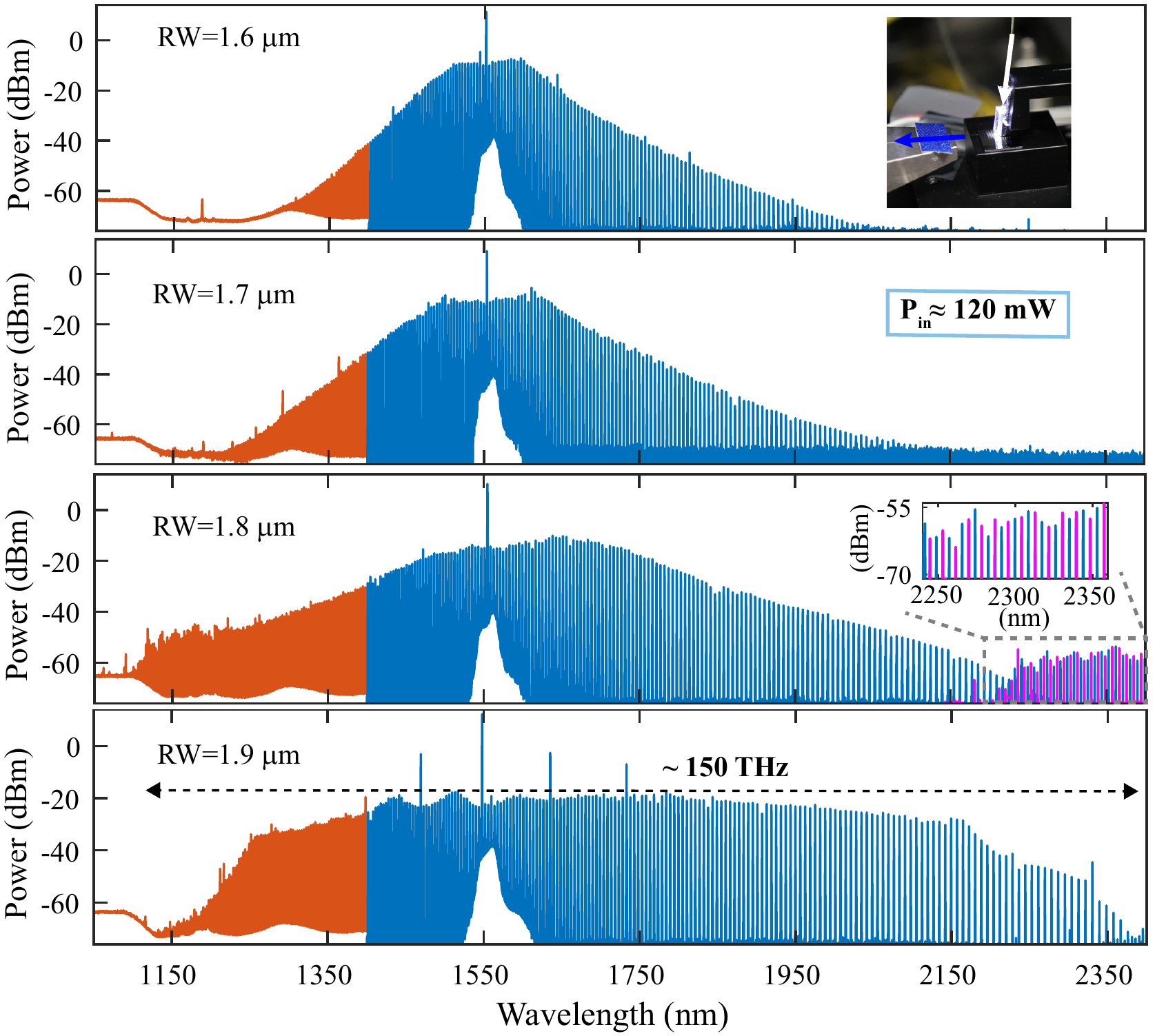}
\caption{Comb spectra for the TE$_{00}$ modes of 36-$\mu$m-radius SiC microrings with varied ring widths and an estimated on-chip power of 120 mW. The orange and light blue colors indicate the data collected from two different optical spectrum analyzers given the wide wavelength range. The inset in the top figure shows the experimental setup for broadband comb measurement: the input which is a continuous-wave signal near the wavelength of 1550 nm is coupled from a fiber V-groove array to the waveguide through a grating coupler, while the broadband output is routed to the polished facet and collected using a lensed fiber (butt coupling). For RW=$1.8\ \mu$m, a secondary mode family appears beyond wavelengths of 2200 nm (see the inset in the third figure from top). The spectral bandwidth corresponding to ring widths of 1.8 $\mu$m and 1.9 $\mu$m is estimated to be larger than 150 THz.}
\label{Fig_TE_comb}
\end{figure}

Having successfully generated combs based on the TM$_{00}$ modes, we proceed to operate on the TE$_{00}$ modes which can potentially offer a much broader comb spectrum. To overcome the limited out-coupling bandwidth from the grating coupler, the experimental configuration has been modified (see the inset of Fig.~4 for the actual setup): we route the output waveguide to a polished chip facet and collect the light using a lensed fiber (butt coupling), which is broadband in nature. On the input side, we still use the grating coupler as it is found to be robust under high input powers. By contrast, the butt coupling often results in burnt chip facet when the incident power is large enough, which is likely caused by the absorption from defect states created during the facet polishing step. In addition, to resolve the broad comb spectrum at the output, two different optical spectrum analyzers (OSAs) are employed, with one covering the wavelength range of 600-1700 nm and the other for 1200-2400 nm. 

Figure 4 plots the comb spectra of the TE$_{00}$ mode for different ring widths of 36-$\mu$m-radius SiC microrings. Consistent with the dispersion profiles provided in Fig.~1(c), we find that the comb bandwidth indeed increases as the anomalous GVD is weakened by the increased waveguide width. In particular, broadband comb states spanning from the wavelength range of 1100 nm to 2400 nm are observed for $\text{RW}=1.8\ \mu$m and $\text{RW}=1.9\ \mu$m with a pump power near 120 mW on chip. To the best of our knowledge, such octave-spanning microcombs are the first demonstration among all SiC-based nanophotonic platforms. While the beat-note measurement indicates the comb states shown in Fig.~4 are MI combs (which partially account for the suppressed dispersive wave for $\text{RW}=1.8\ \mu$m when compared to the simulation data in Fig.~1(d)), attaining such a large bandwidth is a crucial step towards realizing $f$-2$f$ self-referencing. Our next step is to generate coherent soliton microcomb states, with the outstanding challenge to overcome the strong thermo-optic effect as observed in Fig.~3(a) \cite{Li_SiN_octave}. Finally, it is worth mentioning that a careful comparison between the observed comb spectra and the LLE simulation points to a Kerr nonlinear coefficient $n_2=(3.0 \pm 1.0)\times 10^{-19} \text{m}^2/\text{W}$ for our SiC (or an equivalent $\gamma \approx 2.1\ \text{W}^{-1}\text{m}^{-1}$ as used in the LLE simulation in Fig.~1(d)), which is smaller than what have been reported in the literature \cite{Lin_3CSiC_nonlinear, Ou_4HSiC, Vuckovic_4HSiC_MIcomb}. This difference likely results from the material property variation among different SiC manufacturers, as we do observe an even smaller Kerr nonlinear coefficient for semi-insulating 4H-SiC with vanadium doping ($40\%-50\%$ reduction compared to the one used in this work). 
\section{Conclusion}
In conclusion, octave-spanning microcombs are demonstrated on the 4H-SiCOI platform for the first time. This outcome is enabled by efforts on two frontiers: first, optimized nanofabrication leads to compact SiC microrings with intrinsic $Q$s above 1 million; and second, the TE modes are utilized whose dispersion can be tuned in a wide range by varying the waveguide width for a fixed SiC thickness. Our results have paved the way for the chip-scale implementation of the $f$-2$f$ self-referencing in the SiCOI platform, and more generally, lent strong support to the competitiveness of SiC for a variety of nonlinear applications when compared to more mature nonlinear materials such as silicon nitride and aluminum nitride. 

\begin{backmatter}
\bmsection{Funding}
This work was supported by DARPA YFA (D19AP00033) and NSF (2127499). 

\bmsection{Acknowledgments}
The authors would like to thank helpful discussions with Dr.~Gordon Keeler and Dr.~Tianren Fan, and help with the SiC annealing process from Dr.~Xiyuan Lu and Dr.~Kartik Srinivasan at NIST. The OSA data in Fig.~4 was obtained with equipment support from Prof.~Sheng Shen's group at CMU as well as Dr.~Lijun Ma and Dr.~Oliver Slattery at NIST.  
\end{backmatter}


\bibliography{SiC_Ref2}

\begin{thebibliography}{10}
\newcommand{\enquote}[1]{``#1''}

\bibitem{Papp_comb_synthesizer}
D.~T. Spencer, T.~Drake, T.~C. Briles, J.~Stone, L.~C. Sinclair, C.~Fredrick,
  Q.~Li, D.~Westly, B.~R. Ilic, A.~Bluestone, N.~Volet, T.~Komljenovic,
  L.~Chang, S.~H. Lee, D.~Y. Oh, M.-G. Suh, K.~Y. Yang, M.~H.~P. Pfeiffer,
  T.~J. Kippenberg, E.~Norberg, L.~Theogarajan, K.~Vahala, N.~R. Newbury,
  K.~Srinivasan, J.~E. Bowers, S.~A. Diddams, and S.~B. Papp, \enquote{An
  optical-frequency synthesizer using integrated photonics,}
  {\protect\JournalTitle{Nature}} \textbf{557}, 81--85 (2018).

\bibitem{Comb_freq_XK}
J.~Liu, E.~Lucas, A.~S. Raja, J.~He, J.~Riemensberger, R.~N. Wang, M.~Karpov,
  H.~Guo, R.~Bouchand, and T.~J. Kippenberg, \enquote{Photonic microwave
  generation in the {{X}}- and {{K}}-band using integrated soliton microcombs,}
  {\protect\JournalTitle{Nature Photonics}} \textbf{14}, 486--491 (2020).

\bibitem{Wong_SiN_THz}
S.-W. Huang, J.~Yang, S.-H. Yang, M.~Yu, D.-L. Kwong, T.~Zelevinsky,
  M.~Jarrahi, and C.~W. Wong, \enquote{Globally {{stable microresonator Turing
  pattern formation}} for {{coherent high}}-{{power THz radiation
  on}}-{{chip}},} {\protect\JournalTitle{Physical Review X}} \textbf{7}, 041002
  (2017).

\bibitem{Diddams_review_spectrum}
S.~A. Diddams, K.~Vahala, and T.~Udem, \enquote{Optical frequency combs:
  Coherently uniting the electromagnetic spectrum,}
  {\protect\JournalTitle{Science}} \textbf{369}, eaay3676 (2020).

\bibitem{Vahala_comb_imaging}
C.~Bao, M.-G. Suh, and K.~Vahala, \enquote{Microresonator soliton dual-comb
  imaging,} {\protect\JournalTitle{Optica}} \textbf{6}, 1110--1116 (2019).

\bibitem{Diddams_comb_sensing1}
S.~A. Diddams, L.~Hollberg, and V.~Mbele, \enquote{Molecular fingerprinting
  with the resolved modes of a femtosecond laser frequency comb,}
  {\protect\JournalTitle{Nature}} \textbf{445}, 627--630 (2007).

\bibitem{Comb_midIR_spectroscopy}
G.~Ycas, F.~R. Giorgetta, K.~C. Cossel, E.~M. Waxman, E.~Baumann, N.~R.
  Newbury, and I.~Coddington, \enquote{Mid-infrared dual-comb spectroscopy of
  volatile organic compounds across long open-air paths,}
  {\protect\JournalTitle{Optica}} \textbf{6}, 165--168 (2019).

\bibitem{Vahala19_comb_exoplanets}
M.-G. Suh, X.~Yi, Y.-H. Lai, S.~Leifer, I.~S. Grudinin, G.~Vasisht, E.~C.
  Martin, M.~P. Fitzgerald, G.~Doppmann, J.~Wang, D.~Mawet, S.~B. Papp, S.~A.
  Diddams, C.~Beichman, and K.~Vahala, \enquote{Searching for exoplanets using
  a microresonator astrocomb,} {\protect\JournalTitle{Nature Photonics}}
  \textbf{13}, 25--30 (2019).

\bibitem{Vahala_comb_ranging}
M.-G. Suh and K.~J. Vahala, \enquote{Soliton microcomb range measurement,}
  {\protect\JournalTitle{Science}} \textbf{359}, 884--887 (2018).

\bibitem{Kippenberg_comb_ranging}
J.~Riemensberger, A.~Lukashchuk, M.~Karpov, W.~Weng, E.~Lucas, J.~Liu, and
  T.~J. Kippenberg, \enquote{Massively parallel coherent laser ranging using a
  soliton microcomb,} {\protect\JournalTitle{Nature}} \textbf{581}, 164--170
  (2020).

\bibitem{Comb_communication}
P.~{Marin-Palomo}, J.~N. Kemal, M.~Karpov, A.~Kordts, J.~Pfeifle, M.~H.~P.
  Pfeiffer, P.~Trocha, S.~Wolf, V.~Brasch, M.~H. Anderson, R.~Rosenberger,
  K.~Vijayan, W.~Freude, T.~J. Kippenberg, and C.~Koos,
  \enquote{Microresonator-based solitons for massively parallel coherent
  optical communications,} {\protect\JournalTitle{Nature}} \textbf{546},
  274--279 (2017).

\bibitem{Comb_parallel_computation}
J.~Feldmann, N.~Youngblood, M.~Karpov, H.~Gehring, X.~Li, M.~Stappers,
  M.~Le~Gallo, X.~Fu, A.~Lukashchuk, A.~S. Raja, J.~Liu, C.~D. Wright,
  A.~Sebastian, T.~J. Kippenberg, W.~H.~P. Pernice, and H.~Bhaskaran,
  \enquote{Parallel convolutional processing using an integrated photonic
  tensor core,} {\protect\JournalTitle{Nature}} \textbf{589}, 52--58 (2021).

\bibitem{Wong_SiN_quantum}
Z.~Xie, T.~Zhong, S.~Shrestha, X.~Xu, J.~Liang, Y.-X. Gong, J.~C. Bienfang,
  A.~Restelli, J.~H. Shapiro, F.~N.~C. Wong, and C.~Wei~Wong,
  \enquote{Harnessing high-dimensional hyperentanglement through a biphoton
  frequency comb,} {\protect\JournalTitle{Nature Photonics}} \textbf{9},
  536--542 (2015).

\bibitem{Review_quantum_comb}
M.~Kues, C.~Reimer, J.~M. Lukens, W.~J. Munro, A.~M. Weiner, D.~J. Moss, and
  R.~Morandotti, \enquote{Quantum optical microcombs,}
  {\protect\JournalTitle{Nature Photonics}} \textbf{13}, 170--179 (2019).

\bibitem{Diddams_comb_review1}
T.~J. Kippenberg, R.~Holzwarth, and S.~A. Diddams,
  \enquote{Microresonator-{{based optical frequency combs}},}
  {\protect\JournalTitle{Science}} \textbf{332}, 555--559 (2011).

\bibitem{Gaeta_comb_battery}
B.~Stern, X.~Ji, Y.~Okawachi, A.~L. Gaeta, and M.~Lipson,
  \enquote{Battery-operated integrated frequency comb generator,}
  {\protect\JournalTitle{Nature}} \textbf{562}, 401--405 (2018).

\bibitem{Bowers_comb_integration}
C.~Xiang, J.~Liu, J.~Guo, L.~Chang, R.~N. Wang, W.~Weng, J.~Peters, W.~Xie,
  Z.~Zhang, J.~Riemensberger, J.~Selvidge, T.~J. Kippenberg, and J.~E. Bowers,
  \enquote{Laser soliton microcombs heterogeneously integrated on silicon,}
  {\protect\JournalTitle{Science}} \textbf{373}, 99--103 (2021).

\bibitem{Bowers_turnkey_soliton}
B.~Shen, L.~Chang, J.~Liu, H.~Wang, Q.-F. Yang, C.~Xiang, R.~N. Wang, J.~He,
  T.~Liu, W.~Xie, J.~Guo, D.~Kinghorn, L.~Wu, Q.-X. Ji, T.~J. Kippenberg,
  K.~Vahala, and J.~E. Bowers, \enquote{Integrated turnkey soliton microcombs,}
  {\protect\JournalTitle{Nature}} \textbf{582}, 365--369 (2020).

\bibitem{Bowers_comb_laser}
W.~Jin, Q.-F. Yang, L.~Chang, B.~Shen, H.~Wang, M.~A. Leal, L.~Wu, M.~Gao,
  A.~Feshali, M.~Paniccia, K.~J. Vahala, and J.~E. Bowers,
  \enquote{Hertz-linewidth semiconductor lasers using {{CMOS}}-ready
  ultra-high-{{Q}} microresonators,} {\protect\JournalTitle{Nature Photonics}}
  \textbf{15}, 346--353 (2021).

\bibitem{Gaeta_Si_comb}
M.~Yu, Y.~Okawachi, A.~G. Griffith, M.~Lipson, and A.~L. Gaeta,
  \enquote{Mode-locked mid-infrared frequency combs in a silicon
  microresonator,} {\protect\JournalTitle{Optica}} \textbf{3}, 854--860 (2016).

\bibitem{Gaeta_Si_dual_midIR}
M.~Yu, Y.~Okawachi, A.~G. Griffith, N.~Picqu{\'e}, M.~Lipson, and A.~L. Gaeta,
  \enquote{Silicon-chip-based mid-infrared dual-comb spectroscopy,}
  {\protect\JournalTitle{Nature Communications}} \textbf{9}, 1869 (2018).

\bibitem{Vahala_comb_silica}
X.~Yi, Q.-F. Yang, K.~Y. Yang, M.-G. Suh, and K.~Vahala, \enquote{Soliton
  frequency comb at microwave rates in a high-{{Q}} silica microresonator,}
  {\protect\JournalTitle{Optica}} \textbf{2}, 1078--1085 (2015).

\bibitem{Xiao_comb_silica}
X.~Jiang, L.~Shao, S.-X. Zhang, X.~Yi, J.~Wiersig, L.~Wang, Q.~Gong, M.~Lon{\v
  c}ar, L.~Yang, and Y.-F. Xiao, \enquote{Chaos-assisted broadband momentum
  transformation in optical microresonators,} {\protect\JournalTitle{Science}}
  \textbf{358}, 344--347 (2017).

\bibitem{Guo_comb_AlN}
H.~Weng, J.~Liu, A.~A. Afridi, J.~Li, J.~Dai, X.~Ma, Y.~Zhang, Q.~Lu, J.~F.
  Donegan, J.~F. Donegan, W.~Guo, and W.~Guo, \enquote{Directly accessing
  octave-spanning dissipative {{Kerr}} soliton frequency combs in an {{AlN}}
  microresonator,} {\protect\JournalTitle{Photonics Research}} \textbf{9},
  1351--1357 (2021).

\bibitem{Tang_comb_AlN_ref}
X.~Liu, Z.~Gong, A.~W. Bruch, J.~B. Surya, J.~Lu, and H.~X. Tang,
  \enquote{Aluminum nitride nanophotonics for beyond-octave soliton microcomb
  generation and self-referencing,} {\protect\JournalTitle{Nature
  Communications}} \textbf{12}, 5428 (2021).

\bibitem{Gaeta_SiN_octave}
Y.~Okawachi, K.~Saha, J.~S. Levy, Y.~H. Wen, M.~Lipson, and A.~L. Gaeta,
  \enquote{Octave-spanning frequency comb generation in a silicon nitride
  chip,} {\protect\JournalTitle{Optics Letters}} \textbf{36}, 3398--3400
  (2011).

\bibitem{Li_SiN_octave}
Q.~Li, T.~C. Briles, D.~A. Westly, T.~E. Drake, J.~R. Stone, B.~R. Ilic, S.~A.
  Diddams, S.~B. Papp, and K.~Srinivasan, \enquote{Stably accessing
  octave-spanning microresonator frequency combs in the soliton regime,}
  {\protect\JournalTitle{Optica}} \textbf{4}, 193--203 (2017).

\bibitem{Kippenberg_SiN_octave}
M.~H.~P. Pfeiffer, C.~Herkommer, J.~Liu, H.~Guo, M.~Karpov, E.~Lucas,
  M.~Zervas, and T.~J. Kippenberg, \enquote{Octave-spanning dissipative
  {{Kerr}} soliton frequency combs in
  {{Si}}{\textsubscript{3}}{{N}}{\textsubscript{4}} microresonators,}
  {\protect\JournalTitle{Optica}} \textbf{4}, 684--691 (2017).

\bibitem{Loncar_LN_EOM}
M.~Zhang, B.~Buscaino, C.~Wang, A.~{Shams-Ansari}, C.~Reimer, R.~Zhu, J.~M.
  Kahn, and M.~Lon{\v c}ar, \enquote{Broadband electro-optic frequency comb
  generation in a lithium niobate microring resonator,}
  {\protect\JournalTitle{Nature}} \textbf{568}, 373--377 (2019).

\bibitem{Lin_LN_soliton}
Y.~He, Q.-F. Yang, J.~Ling, R.~Luo, H.~Liang, M.~Li, B.~Shen, H.~Wang,
  K.~Vahala, and Q.~Lin, \enquote{Self-starting bi-chromatic
  {{LiNbO}}{\textsubscript{3}} soliton microcomb,}
  {\protect\JournalTitle{Optica}} \textbf{6}, 1138--1144 (2019).

\bibitem{Tang_LN_comb}
Z.~Gong, X.~Liu, Y.~Xu, and H.~X. Tang, \enquote{Near-octave lithium niobate
  soliton microcomb,} {\protect\JournalTitle{Optica}} \textbf{7}, 1275--1278
  (2020).

\bibitem{Pu_AlGAAs_comb}
M.~Pu, L.~Ottaviano, E.~Semenova, and K.~Yvind, \enquote{Efficient frequency
  comb generation in {{AlGaAs}}-on-insulator,} {\protect\JournalTitle{Optica}}
  \textbf{3}, 823--826 (2016).

\bibitem{Bowers_comb_AlGaAs}
L.~Chang, W.~Xie, H.~Shu, Q.-F. Yang, B.~Shen, A.~Boes, J.~D. Peters, W.~Jin,
  C.~Xiang, S.~Liu, G.~Moille, S.-P. Yu, X.~Wang, K.~Srinivasan, S.~B. Papp,
  K.~Vahala, and J.~E. Bowers, \enquote{Ultra-efficient frequency comb
  generation in {{AlGaAs}}-on-insulator microresonators,}
  {\protect\JournalTitle{Nature Communications}} \textbf{11}, 1331 (2020).

\bibitem{Awschalom_SiC_qubit}
C.~P. Anderson, A.~Bourassa, K.~C. Miao, G.~Wolfowicz, P.~J. Mintun, A.~L.
  Crook, H.~Abe, J.~U. Hassan, N.~T. Son, T.~Ohshima, and D.~D. Awschalom,
  \enquote{Electrical and optical control of single spins integrated in
  scalable semiconductor devices,} {\protect\JournalTitle{Science}}
  \textbf{366}, 1225--1230 (2019).

\bibitem{Awschalom_SiC2}
G.~Wolfowicz, C.~P. Anderson, B.~Diler, O.~G. Poluektov, F.~J. Heremans, and
  D.~D. Awschalom, \enquote{Vanadium spin qubits as telecom quantum emitters in
  silicon carbide,} {\protect\JournalTitle{Science Advances}} \textbf{6},
  eaaz1192 (2020).

\bibitem{Vuckovic_SiC_review}
D.~M. Lukin, M.~A. Guidry, and J.~Vu{\v c}kovi{\'c}, \enquote{Integrated
  {{quantum photonics}} with {{silicon carbide}}: challenges and
  {{prospects}},} {\protect\JournalTitle{PRX Quantum}} \textbf{1}, 020102
  (2020).

\bibitem{SiC_nonlinear_coeff}
H.~Sato, M.~Abe, I.~Shoji, J.~Suda, and T.~Kondo, \enquote{Accurate
  measurements of second-order nonlinear optical coefficients of {{6H}} and
  {{4H}} silicon carbide,} {\protect\JournalTitle{JOSA B}} \textbf{26},
  1892--1896 (2009).

\bibitem{Lin_3CSiC_nonlinear}
X.~Lu, J.~Y. Lee, S.~Rogers, and Q.~Lin, \enquote{Optical {{Kerr}} nonlinearity
  in a high-{{Q}} silicon carbide microresonator,}
  {\protect\JournalTitle{Optics Express}} \textbf{22}, 30826 (2014).

\bibitem{Gaeta_4HSiC_nonlinear}
J.~Cardenas, M.~Yu, Y.~Okawachi, C.~B. Poitras, R.~K.~W. Lau, A.~Dutt, A.~L.
  Gaeta, and M.~Lipson, \enquote{Optical nonlinearities in high-confinement
  silicon carbide waveguides,} {\protect\JournalTitle{Optics Letters}}
  \textbf{40}, 4138--4141 (2015).

\bibitem{Lin_3CSiC}
X.~Lu, J.~Y. Lee, P.~X.-L. Feng, and Q.~Lin, \enquote{Silicon carbide microdisk
  resonator,} {\protect\JournalTitle{Optics Letters}} \textbf{38}, 1304--1306
  (2013).

\bibitem{Adibi_3CSiC}
T.~Fan, H.~Moradinejad, X.~Wu, A.~A. Eftekhar, and A.~Adibi,
  \enquote{High-{{Q}} integrated photonic microresonators on
  {{3C}}-{{SiC}}-on-insulator ({{SiCOI}}) platform,}
  {\protect\JournalTitle{Optics Express}} \textbf{26}, 25814--25826 (2018).

\bibitem{Adibi_3CSiC_Q}
T.~Fan, X.~Wu, A.~A. Eftekhar, M.~Bosi, H.~Moradinejad, E.~V. Woods, and
  A.~Adibi, \enquote{High-quality integrated microdisk resonators in the
  visible-to-near-infrared wavelength range on a {{3C}}-silicon
  carbide-on-insulator platform,} {\protect\JournalTitle{Optics Letters}}
  \textbf{45}, 153 (2020).

\bibitem{Ou_4HSiC}
Y.~Zheng, M.~Pu, A.~Yi, X.~Ou, and H.~Ou, \enquote{{{4H}}-{{SiC}} microring
  resonators for nonlinear integrated photonics,} {\protect\JournalTitle{Optics
  Letters}} \textbf{44}, 5784 (2019).

\bibitem{Noda_4HSiC_PhC}
B.-S. Song, T.~Asano, S.~Jeon, H.~Kim, C.~Chen, D.~D. Kang, and S.~Noda,
  \enquote{Ultrahigh-{{Q}} photonic crystal nanocavities based on {{4H}}
  silicon carbide,} {\protect\JournalTitle{Optica}} \textbf{6}, 991 (2019).

\bibitem{Vuckovic_4HSiC_nphoton}
D.~M. Lukin, C.~Dory, M.~A. Guidry, K.~Y. Yang, S.~D. Mishra, R.~Trivedi,
  M.~Radulaski, S.~Sun, D.~Vercruysse, G.~H. Ahn, and J.~Vu{\v c}kovi{\'c},
  \enquote{{{4H}}-silicon-carbide-on-insulator for integrated quantum and
  nonlinear photonics,} {\protect\JournalTitle{Nature Photonics}} \textbf{14},
  330--334 (2020).

\bibitem{Ou_4HSiC_combQ}
C.~Wang, Z.~Fang, A.~Yi, B.~Yang, Z.~Wang, L.~Zhou, C.~Shen, Y.~Zhu, Y.~Zhou,
  R.~Bao, Z.~Li, Y.~Chen, K.~Huang, J.~Zhang, Y.~Cheng, and X.~Ou,
  \enquote{High-{{Q}} microresonators on {{4H}}-silicon-carbide-on-insulator
  platform for nonlinear photonics,} {\protect\JournalTitle{Light: Science \&
  Applications}} \textbf{10}, 139 (2021).

\bibitem{Vuckovic_4HSiC_soliton}
M.~A. Guidry, D.~M. Lukin, K.~Y. Yang, R.~Trivedi, and J.~Vu{\v c}kovi{\'c},
  \enquote{Quantum optics of soliton microcombs,}
  {\protect\JournalTitle{arXiv:2103.10517 [physics, physics:quant-ph]}}
  (2021).

\bibitem{Awschalom_SiC_PhC}
A.~L. Crook, C.~P. Anderson, K.~C. Miao, A.~Bourassa, H.~Lee, S.~L. Bayliss,
  D.~O. Bracher, X.~Zhang, H.~Abe, T.~Ohshima, E.~L. Hu, and D.~D. Awschalom,
  \enquote{Purcell {{enhancement}} of a {{single silicon carbide color center}}
  with {{coherent spin control}},} {\protect\JournalTitle{Nano Letters}}
  \textbf{20}, 3427--3434 (2020).

\bibitem{Vuckovic_4HSiC_MIcomb}
M.~A. Guidry, K.~Y. Yang, D.~M. Lukin, A.~Markosyan, J.~Yang, M.~M. Fejer, and
  J.~Vu{\v c}kovi{\'c}, \enquote{Optical parametric oscillation in silicon
  carbide nanophotonics,} {\protect\JournalTitle{Optica}} \textbf{7}, 1139
  (2020).

\bibitem{Lin_3CSiC_double}
X.~Lu, J.~Y. Lee, S.~D. Rogers, and Q.~Lin, \enquote{Silicon carbide
  double-microdisk resonator,} {\protect\JournalTitle{Optics Letters}}
  \textbf{44}, 4295 (2019).

\bibitem{Diddams_comb_selfref}
A.~Bartels, D.~Heinecke, and S.~A. Diddams, \enquote{10-{{GHz
  self}}-{{referenced optical frequency comb}},}
  {\protect\JournalTitle{Science}} \textbf{326}, 681--681 (2009).

\bibitem{Gaeta_LN_selfref}
Y.~Okawachi, M.~Yu, B.~Desiatov, B.~Y. Kim, T.~Hansson, M.~Lon{\v c}ar, and
  A.~L. Gaeta, \enquote{Chip-based self-referencing using integrated lithium
  niobate waveguides,} {\protect\JournalTitle{Optica}} \textbf{7}, 702--707
  (2020).

\bibitem{Kim_comb_dispersion}
S.~Kim, K.~Han, C.~Wang, J.~A. {Jaramillo-Villegas}, X.~Xue, C.~Bao, Y.~Xuan,
  D.~E. Leaird, A.~M. Weiner, and M.~Qi, \enquote{Dispersion engineering and
  frequency comb generation in thin silicon nitride concentric
  microresonators,} {\protect\JournalTitle{Nature Communications}} \textbf{8},
  372 (2017).

\bibitem{Li_FWMBS}
Q.~Li, M.~Davan{\c c}o, and K.~Srinivasan, \enquote{Efficient and low-noise
  single-photon-level frequency conversion interfaces using silicon
  nanophotonics,} {\protect\JournalTitle{Nature Photonics}} \textbf{10},
  406--414 (2016).

\bibitem{Kippenberg_soliton}
T.~Herr, V.~Brasch, J.~D. Jost, C.~Y. Wang, N.~M. Kondratiev, M.~L. Gorodetsky,
  and T.~J. Kippenberg, \enquote{Temporal solitons in optical microresonators,}
  {\protect\JournalTitle{Nature Photonics}} \textbf{8}, 145--152 (2014).

\bibitem{Kippenberg_DW}
V.~Brasch, M.~Geiselmann, T.~Herr, G.~Lihachev, M.~H.~P. Pfeiffer, M.~L.
  Gorodetsky, and T.~J. Kippenberg, \enquote{Photonic chip-based optical
  frequency comb using soliton cherenkov radiation,}
  {\protect\JournalTitle{Science}} \textbf{351}, 357--360 (2016).

\bibitem{Coen_LLE}
S.~Coen, H.~G. Randle, T.~Sylvestre, and M.~Erkintalo, \enquote{Modeling of
  octave-spanning kerr frequency combs using a generalized mean-field
  lugiato-lefever model,} {\protect\JournalTitle{Opt. Lett.}} \textbf{38},
  37--39 (2013).

\bibitem{Papp_soliton_crystal}
D.~C. Cole, E.~S. Lamb, P.~Del'Haye, S.~A. Diddams, and S.~B. Papp,
  \enquote{Soliton crystals in {{Kerr}} resonators,}
  {\protect\JournalTitle{Nature Photonics}} \textbf{11}, 671--676 (2017).

\end{thebibliography}

\end{document}